\documentclass[aip,rsi,12pt]{revtex4-1}
\usepackage{units}
\usepackage{graphicx}

\begin{document}

\title{A dark-field microscope for background-free detection of resonance fluorescence from single semiconductor quantum dots operating in a set-and-forget mode} %Title of paper



\author{Andreas V. Kuhlmann}
\email[]{andreas.kuhlmann@unibas.ch}
\affiliation{Department of Physics, University of Basel, Klingelbergstrasse 82, CH-4056 Basel, Switzerland}

\author{Julien Houel}
\affiliation{Department of Physics, University of Basel, Klingelbergstrasse 82, CH-4056 Basel, Switzerland}

\author{Daniel Brunner}
\affiliation{Instituto de F\'{i}sica Interdisciplinar y Sistemas Complejos, IFISC (CSIC-UIB), Campus Universitat Illes Balears, E-07122 Palma de Mallorca, Spain}

\author{Arne Ludwig}
\affiliation{Department of Physics, University of Basel, Klingelbergstrasse 82, CH-4056 Basel, Switzerland}
\affiliation{Lehrstuhl f\"{u}r Angewandte Festk\"{o}rperphysik, Ruhr-Universit\"{a}t Bochum, D-44780 Bochum, Germany}

\author{Dirk Reuter}
\affiliation{Lehrstuhl f\"{u}r Angewandte Festk\"{o}rperphysik, Ruhr-Universit\"{a}t Bochum, D-44780 Bochum, Germany}
\affiliation{Department Physik, Universit\"{a}t Paderborn, Warburger Strasse 100, D-33098 Paderborn, Germany}

\author{Andreas D. Wieck}
\affiliation{Lehrstuhl f\"{u}r Angewandte Festk\"{o}rperphysik, Ruhr-Universit\"{a}t Bochum, D-44780 Bochum, Germany}

\author{Richard J. Warburton}
\affiliation{Department of Physics, University of Basel, Klingelbergstrasse 82, CH-4056 Basel, Switzerland}
%\thanks{}


\date{\today}

\begin{abstract}
Optically active quantum dots, for instance self-assembled InGaAs quantum dots, are potentially excellent single photon sources. The fidelity of the single photons is much improved using resonant rather than non-resonant excitation. With resonant excitation, the challenge is to distinguish between resonance fluorescence and scattered laser light. We have met this challenge by creating a polarization-based dark-field microscope to measure the resonance fluorescence from a single quantum dot at low temperature. We achieve a suppression of the scattered laser exceeding a factor of $10^{7}$ and background-free detection of resonance fluorescence. The same optical setup operates over the entire quantum dot emission range ($\unit[920-980]{\rm nm}$) and also in high magnetic fields. The major development is the outstanding long-term stability: once the dark-field point has been established, the microscope operates for days without alignment. The mechanical and optical design of the microscope is presented, as well as exemplary resonance fluorescence spectroscopy results on individual quantum dots to underline the microscope's excellent performance.
\end{abstract}


\maketitle %\maketitle must follow title, authors, abstract and \pacs

\section{Introduction}
%\label{}
Semiconductor quantum dots, in particular self-assembled InGaAs quantum dots, are very attractive as the building blocks for quantum light sources\cite{Shields2007} and spin qubits\cite{Warburton2013}. Self-assembled InGaAs quantum dots (operating at wavelengths around 950 nm at low temperature) exploit technologically advanced GaAs heterostructures and have become the workhorse system in the field. It is hugely advantageous to explore the physics using resonant rather than non-resonant laser excitation. On the one hand, non-resonant excitation introduces sources of noise resulting in exciton and spin dephasing\cite{Kuhlmann2013}. On the other hand, resonant (but not non-resonant) excitation allows a  spin to be initialized\cite{Atature2006,Gerardot2008}, manipulated\cite{Press2008} and read-out\cite{Vamivakas2010} optically. Resonant excitation, i.e.\ coherent laser spectroscopy, on single InGaAs/GaAs quantum dots was first developed with differential transmission detection\cite{Hogele2004}, using Stark-shift modulation of the transitions energy along with lock-in detection for noise rejection \cite{Al'en2003}. The detection scheme exploits an interference between the laser field and the field associated with coherently scattered photons\cite{Karrai2003}: it provides a sensitive detection scheme but does not provide direct access to the resonance fluorescence, the single photons scattered or emitted by the quantum dot. These photons are crucial to develop a high-fidelity single photon source and, further afield, in developing a quantum dot-based quantum network with applications in quantum communication\cite{Kimble2008}.

Recently, the resonance fluorescence of a semiconductor quantum dot\cite{Muller2007,Ates2009,Nguyen2011,Ulhaq2012,Vamivakas2009,Vamivakas2010a,Yilmaz2010,Matthiesen2012,Houel2012,Gao2012,Kuhlmann2013} has been observed. The challenge experimentally is to distinguish quantum dot-scattered light from scattered laser light. With non-resonant excitation, this separation is trivial to achieve on account of the widely different wavelengths. With resonant excitation, this scheme fails. One scheme for the detection of resonance fluorescence exploits the different wave vectors of the laser light and the resonance fluorescence\cite{Ates2009,Muller2007,Nguyen2011,Ulhaq2012}. This is very much in the spirit of the original ensemble experiments in atomic physics in which resonance fluorescence was detected in a direction orthogonal to the carefully defined propagation direction of the laser\cite{Schuda1974,Wu1975}. In a semiconductor context, one implementation of this scheme involves coupling laser light to a waveguide containing quantum dots with edge illumination, detecting the resonance fluorescence in the orthogonal vertical direction\cite{Muller2007,Ates2009,Ulhaq2012}. Another scheme exploits a further property of light: its polarization. The idea is to operate in the dark-field as defined by the polarization: the laser and the detection are defined to have orthogonal polarization states. Provided laser scattering preserves the polarization, the crossed polarizer configuration ensures that scattered laser light is prevented from entering the detection mode. Success has been achieved using crossed linear polarizations\cite{Vamivakas2009,Vamivakas2010a,Yilmaz2010,Matthiesen2012,Houel2012,Gao2012,Kuhlmann2013}.

In our experiments, we have pursued the polarization-based dark-field technique as, first, it doesn't require a specially fabricated waveguide and second, space limitations in the bore of a superconducting magnet limit the possibilities for efficient edge illumination. It is clear that achieving sufficient laser rejection based on polarization requires both high quality polarizing optics and exquisite angular control. Our first experiments achieved success but only for times of a few minutes after which the dark-field setting had to be re-optimized. This is likely to be a common problem. Here, we present both the mechanical and optical design of a dark-field microscope for resonance fluorescence experiments on a quantum dot. All the figures of merit are excellent, state-of-the art or better: polarization filtering allows us to suppress the excitation laser in the detection beam path by up to 8 orders of magnitude; in combination with a reasonably high light collection efficiency, resonance fluorescence can be measured with a signal-to-background ratio exceeding $10^4:1$. The property we emphasize however is {\em stability}. The long-term stability is such that the microscope can be operated for many days in a set-and-forget mode.

Our own motivation for developing the dark-field microscope was to push forward a research programme on single self-assembled quantum dots. However, we stress that our dark-field microscope is not limited to this field. It will be a perfect tool in the exploration of other quantum emitters for instance colour centres in diamond, single molecules and colloidal quantum dots.

\section{Description of the dark-field microscope}
The design of the dark-field microscope makes no particular demands on the sample although a flat, smooth surface is best. Once the optics' wavelength range is adapted to the emission range, the dark field concept operates equally well with the sample at room temperature or at low temperature. Here, as an example of a two-level system in the solid state, we study self-assembled InGaAs quantum dots emitting at wavelengths around $\unit[950]{nm}$ at low temperature. The microscope combines both high spatial resolution, implemented by a confocal set-up, and dark-field performance. It is designed to allow background-free detection of resonance fluorescence while operating in a set-and-forget mode.  

\subsection{Dark-field concept}
The excitation and detection beams both follow the main axis of the microscope. Thus, laser light back-reflected at the sample has to be suppressed. Apart from its spatial mode, monochromatic laser light is characterized by two distinct features: its frequency and its state of polarization. Laser light cannot be distinguished from the quantum dot emission in frequency as it is a resonant scattering process. However, the state of polarization allows a discrimination to be made between laser and quantum dot photons. The light excitation and detection polarization states have to be orthogonal, here linear s and linear p. 

\begin{figure}[htbp]  
\center
\includegraphics[width=0.65\linewidth]{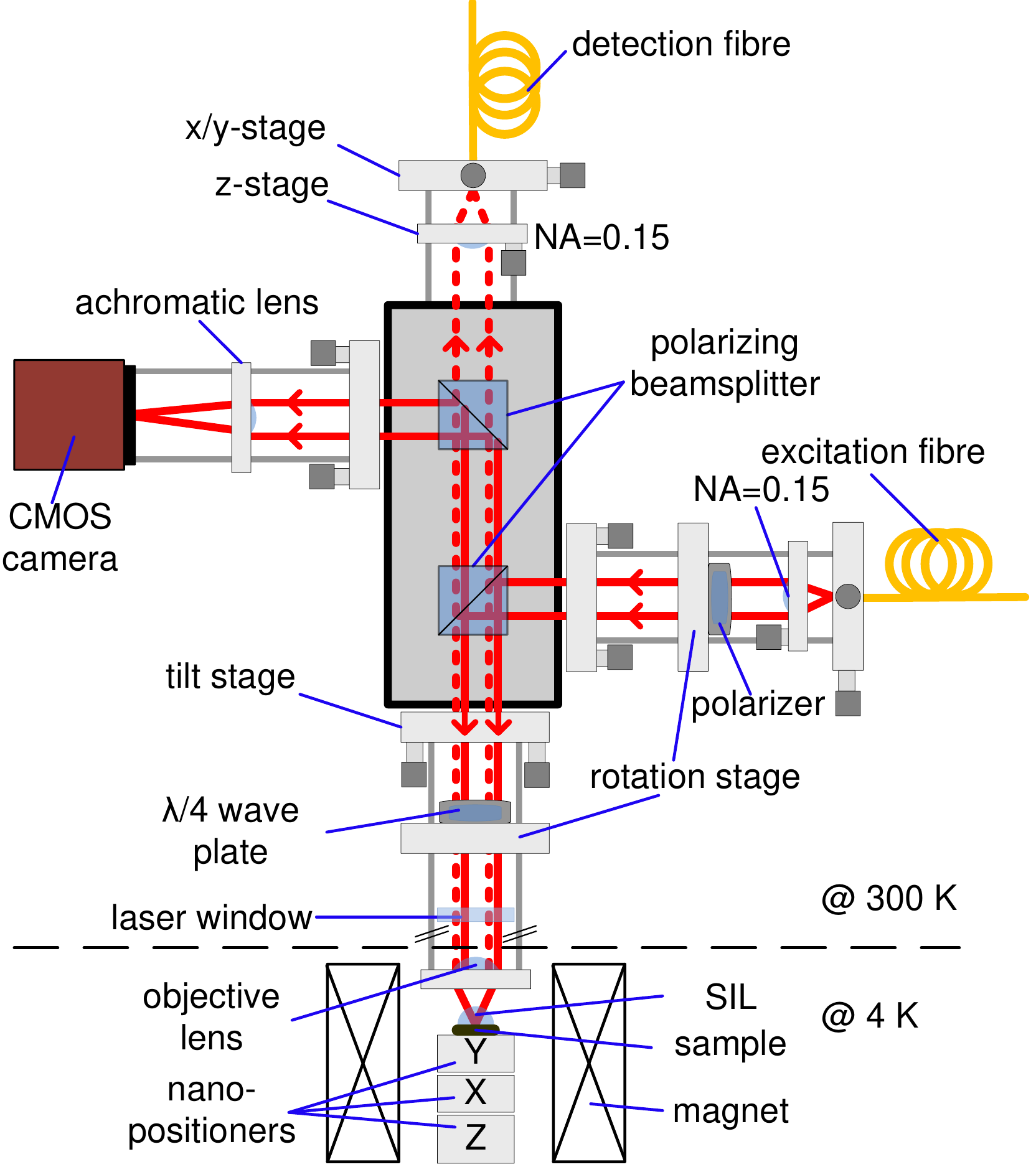}
\caption{Microscope set-up for resonance fluorescence experiments on a single InGaAs quantum dot. Two-level systems implemented in different materials can be studied both at room temperature and at low temperature. Here, a set-up to probe semiconductor quantum dots is shown. The sample with a hemispherical solid immersion lens (SIL) and the objective lens are located inside a bath cryostat, the rest of the microscope remains at room temperature. Optical access is provided by a sealed laser window. The microscope design: three modules, the lower horizontal microscope arm, the vertical arm and the upper horizontal arm are fixed to a central cage containing two polarizing beam splitters (PBSs). The excitation laser is injected via the lower horizontal arm; the vertical arm is used for detection; and the upper horizontal arm for imaging the sample surface. Optical fibres connect the microscope to lasers and detectors mounted on an adjacent optical table. Laser suppression is implemented by means of orthogonal excitation/collection polarization states: the linear polarizer sets the laser polarization to s, matching the lower PBS; the quarter-wave plate controls the state of polarization; and the PBSs reject the s-polarized back-reflected laser light. Solid lines indicate s-polarization, dashed lines p-polarization.}
\label{fig:OpticsSchematic}
\end{figure}

Laser light suppression is implemented by means of orthogonal excitation/collection polarization states, in our case by two polarizing beam splitters (PBS), one linear polarizer and a quarter-wave plate. Their spatial arrangement is shown in Fig.\ \ref{fig:OpticsSchematic}. The purpose of the PBSs is to reject back-scattered laser light; the linear polarizer and quarter-wave plate define and control the state of light polarization. In this scheme, the PBSs define linear s-polarization and linear p-polarization for excitation and detection, respectively. The linear polarizer sets the polarization of the laser light to s-polarization before striking the PBS, the quarter-wave plate controls the polarization thereafter. In particular, the quarter-wave plate allows for a compensation should an elliptically be inadvertently induced. The back-scattered s-polarized laser light is reflected by both the first and second PBS by $\unit[90]{^{\circ}}$ such that the s-polarization is highly suppressed in transmission. The p-polarized component of the quantum dot emission, however, is transmitted and can be detected.

\subsection{Dark-field microscope design}
Experiments on single semiconductor quantum dots typically require low temperatures. The dark-field microscope is therefore integrated into a free-beam microscope system developed for low temperature experiments. The microscope optics apart from the objective lens remain under ambient conditions, as shown in Fig.\ \ref{fig:OpticsSchematic}. The construction frame for the microscope ``head" is a $\unit[30]{mm}$ cage system that allows a modular design: the lower horizontal microscope ``arm'' provides the excitation laser, the vertical arm is used for light detection and the upper horizontal arm to image the sample surface. Each module is attached to a central cage, hosting the PBSs.  
  
The lower horizontal microscope arm provides a link between remote excitation sources and the microscope. Its output is a well collimated beam of coherent laser light, precisely controlled in linear polarization, and used to excite a single quantum dot resonantly. A single mode (SM) fibre (FONT Canada SM fibre $\mbox{NA} = 0.12$, mode field diameter (MDF) $\unit[5.2]{ \mu m}$) interconnects the microscope and the excitation laser (Toptica DL pro 940). By adjusting an x/y-translation stage (Thorlabs CP1XY), the fibre core can be centered on the optical axis defined by the collimator (Thorlabs C280TME-B $\mbox{NA} = 0.15$, $\mbox{f} = \unit[18.4]{mm}$) which is mounted in a z-translation stage (Thorlabs SM1Z). Aspheric lenses are used to collimate/focus the laser beam, as they provide diffraction limited performance for monochromatic applications. A metallic nanoparticle linear film polarizer (Thorlabs LPVIS050-MP) mounted on a rotary stepper positioner (attocube ANR240) polarizes the excitation laser linearly and additionally allows the axis of linear polarization to be precisely controlled. The piezo-driven rotary stepper positioner provides both $\unit[360]{^{\circ}}$ endless rotation and a step size as small as $\unit[1]{\rm m^{\circ}}$. Furthermore, after aligning the polarizer position by means of the control electronics (attocube ANC300), the piezos are grounded and their position is locked, providing outstanding long-term stability. The four cage rods of the excitation arm are connected to a tilt stage (Thorlabs KC1-T/M) which is attached to the central beam splitter cage and allows for a compensation of any angular displacement of the beam.     

The vertical microscope arm is designed to collect light efficiently with a confocal rejection of any stray light. This relies on coupling into a SM fibre (FONT Canada SM fibre $\mbox{NA} = 0.12$, MDF $\unit[5.2]{\mu m}$) which interconnects the microscope and the detectors. The same optical and opto-mechanical components as for the light collimation unit of the horizontal arm are used. The vertical arm is assembled directly on to the PBS cage. The lower tilt stage allows to correct for a misalignment with respect to the optical axis of the objective lens (Thorlabs 352330-B $\mbox{NA} = 0.68$, $\mbox{f} = \unit[3.1]{mm}$).

The upper horizontal microscope arm provides the possibility of monitoring the objective lens focal plane, i.e.\ the sample surface. An achromatic lens (Thorlabs AC254-150-B-ML, $\mbox{f} = \unit[150]{mm}$) focuses light onto the chip of a complementary metal oxide semiconductor (CMOS) camera (Allied Vision Technologies Guppy F-503B), resulting in a magnified image (magnification of 48) of the sample surface. Again a tilt stage allows angular control of the optical axis.

All modules of the microscope are attached to a central cage made from a solid piece of aluminum. It provides stability to the microscope and at the same time hosts two PBSs (B.\ Halle \& Nachfolger PTW 2.10), crucial to implement the polarization filtering. The PBSs allow beam splitting sensitive to the polarization of the incident beam. Two right angle prisms made of flint glass are cemented together to form a cube. A dielectric beam-splitter coating which is deposited on one of the prisms provides a close to unity transmission for p-polarized and close to zero transmission for s-polarized light.

A quarter-wave plate (B.\ Halle \& Nachfolger RZQ 4.10) is mounted beneath the PBSs on a second piezo rotary stage. (Note that the quarter-wave plate behaves as a half-wave plate for the reflected laser light as the laser beam passes it twice.) On the one hand, it is useful during the setup procedure to misalign the quarter-wave plate deliberately and allow some reflected laser light into the detection arm. On the other hand, the quarter-wave plate represents an extra degree of freedom and it turns out that this is crucial: it compensates for any distortion from linear to elliptical polarization in the two polarization states. It is not exactly clear where these small distortions arise, but they are probably related to a birefringence of the sample (GaAs with thin metal layer), objective lens or the cryostat window. The quarter-wave plate used here is a zero order wave plate designed for $\unit[946]{nm}$ and was chosen because, first, it is less temperature sensitive than the multi-order counterparts; and, second, its performance at these particular wavelengths $\unit[950\pm20]{nm}$ surpasses the performance of achromatic wave plates. Again a crucial point for the long-term behavior of the dark-field microscope is that the quarter-wave plate is mounted on a piezo positioner, as for the linear polarizer.

The microscope is inserted into a $\unit[2]{\rm inch}$ bore stainless steel bore, evacuated, and filled with $\unit[\sim 25]{\rm mbar}$ He gas (exchange gas) at room temperature. The tube is then slowly inserted into a He bath cryostat equipped with a $\unit[9]{\rm T}$ superconducting solenoid. The optics at $\unit[300]{\rm K}$ are possibly subject to thermal drift but these are minimized by working in a $\pm \unit[1]{^\circ \rm C}$ temperature stabilized laboratory. 

\subsection{Dark-field microscope alignment}
The microscope operates in both confocal and dark-field modes. For confocal performance, the excitation and collection beams must be concentric and parallel to the optical axis of the objective lens. While monitoring the focal spots on the sample surface, the tilt stages are aligned in order to superimpose the focal spots. The $z$-position of the sample relative to the objective focal plane is adjusted by moving the sample with nanometer precision. During this alignment step, laser light is also coupled into the fibre of the vertical microscope arm. Subsequently, once the confocal condition has been achieved, the linear polarizer and quarter-wave plate are aligned to suppress the back-reflected laser light. In a first step, the linear polarizer is aligned to define the polarization of the laser to s. One approach is to monitor the transmitted signal at the PBS as the polarizer is moved. A minimum in transmission is required. In a second step, the quarter-wave plate is rotated for optimum laser suppression. We find that iterative fine tuning of the polarizer and quarter-wave plate angles enhances the rejection further. Piezo-electronics allow remote control of both the angle of the linear polarizer and the quarter-wave plate. Once the angles are set, the piezos are grounded. 

\section{Dark-field microscope performance}
The performance of the dark-field microscope is characterized under real, experimental conditions: the laser is focused on a quantum dot sample in a low temperature experiment. 

\subsection{Quantum dot sample}
The InGaAs quantum dots are grown by molecular beam epitaxy utilizing a strain-driven self-assembly process and are embedded in a Schottky diode\cite{Drexler1994,Warburton2000}. They are separated from an n$^+$ back contact by a $\unit[25]{\rm nm}$ thick GaAs tunnel barrier. On top of the quantum dots is a capping layer of thickness $\unit[150]{\rm nm}$, followed by a blocking barrier, an AlAs/GaAs superlattice of thickness $\unit[272]{nm}$. The samples are processed with Ohmic contacts to the back contact, grounded in the experiment, and with a semi-transparent gate electrode on the surface (3/7 nm Ti/Au) to which a gate voltage $V_g$ is applied. The number of carriers confined to the quantum dot can be precisely controlled by the applied voltage, allowing the different charged excitons to be addressed. Detuning of the exciton energy with respect to the constant laser frequency is achieved by sweeping $V_g$ on account of the dc Stark effect. The laser spectroscopy is carried out at $\unit[4.2]{\rm K}$ by focusing a $\unit[1]{\rm MHz}$ linewidth laser to a $\unit[0.5]{\mu \rm m}$ spot on the sample surface. A ZrO$_2$ solid immersion lens is mounted directly on top of the sample in order to enhance the collection efficiency and to reduce the spot size\cite{Gerardot2007}. The signal is recorded with a silicon avalanche photodiode (Excelitas SPCM-AQRH-16, photon detection efficiency at $\unit[950]{nm} \sim \unit[25]{\%}$, dark count rate 14 Hz) in photon counting mode.

\subsection{Laser suppression and long-term stability}
\begin{figure}[b]
\includegraphics{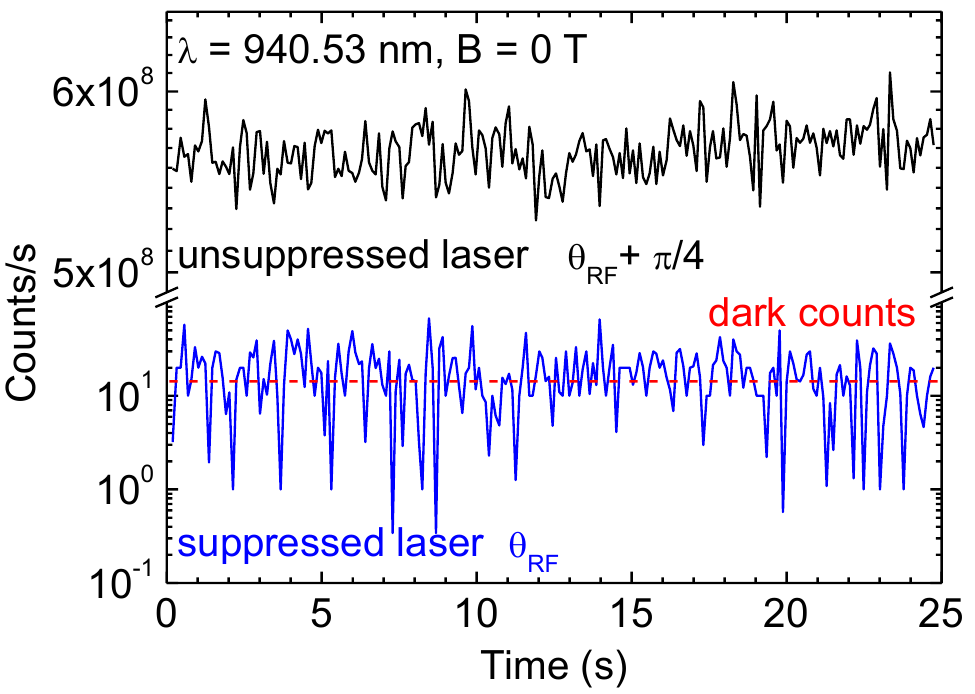}
\caption{Laser suppression. Reflected laser light is monitored at two different angles $\theta$ of the quarter-wave plate. At $\theta = \theta_{\rm RF}$ (blue line) the laser is optimally filtered, at $\theta_{\rm RF}+\pi/4$ (black line) it is optimally transmitted. The count rate decreases from $\unit[580]{\rm MHz}$ to $\unit[4]{\rm Hz}$ (corrected for dark counts), corresponding to a laser suppression exceeding 8 orders of magnitude. A silicon avalanche photodiode in photon counting mode with a dark count rate of \unit[14]{\rm Hz} (dashed red line) is used to detect the laser light reflected from a quantum dot sample (GaAs plus thin metal layer, reflectivity $\sim 50$\%). Integration time per point $\unit[0.1]{\rm s}$.}
\label{figure2}
\end{figure}
In order to observe resonance fluorescence with a high signal-to-background ratio the microscope's laser suppression has to be high. The laser rejection can be determined by rotating the quarter-wave plate, switching between laser rejection maximally on and maximally off. The back-reflected laser light intensity depends periodically on the quarter-wave plate angle with a period of $\pi/4$. A laser suppression exceeding $10^8$, corresponding to an optical density (OD) of 8 is achieved. (The OD is defined as ${\rm OD} = -\log(1/T)$ with transmission $T$.) Fig.\ \ref{figure2} shows a time trace of the detected laser light with and without laser rejection. An initial count rate of $\unit[580]{\rm MHz}$ is reduced to $\unit[4]{\rm Hz}$ by switching on the suppression.    
 
\begin{figure}[b]
 \includegraphics{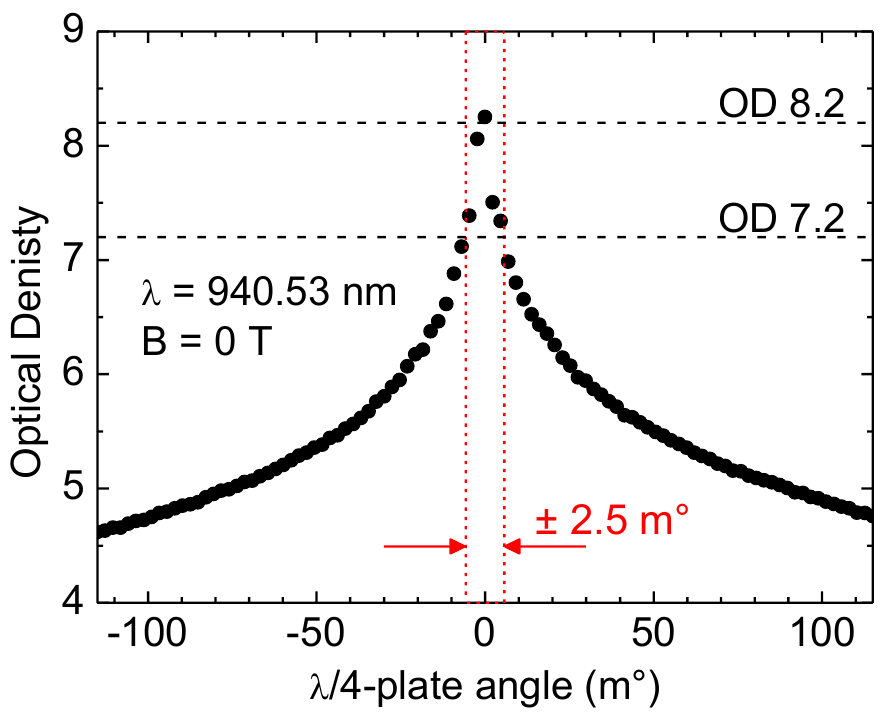}
 \caption{Sensitivity of the laser suppression to the quarter-wave plate angle. The laser light reflected at the quantum dot sample (GaAs plus thin metal layer) is recorded by a silicon avalanche photodiode in photon counting mode as the quarter-wave plate angle is varied, and the corresponding optical density is calculated. At the angle of optimum laser rejection (OD $>$ 8) a change in angle of only $\unit[2.5]{\rm m^{\circ}}$ causes the OD to decrease by one order of magnitude.}
 \label{figure3}
 \end{figure}

The effort to align the dark-field microscope is low. However, how stable is the alignment? Fig.\ \ref{figure3} shows how the optical density depends on the quarter-wave plate angle: it is an extremely sensitive dependence. A change as small as a few $\rm m^{\circ}$ can worsen the rejection by one order of magnitude. On the one hand, it emphasizes the need for a $\rm m^{\circ}$ positioning resolution and on the other hand, the need for an extreme mechanical and thermal stability to achieve good long-term dark-field performance. 
 
\begin{figure}[t]
 \includegraphics{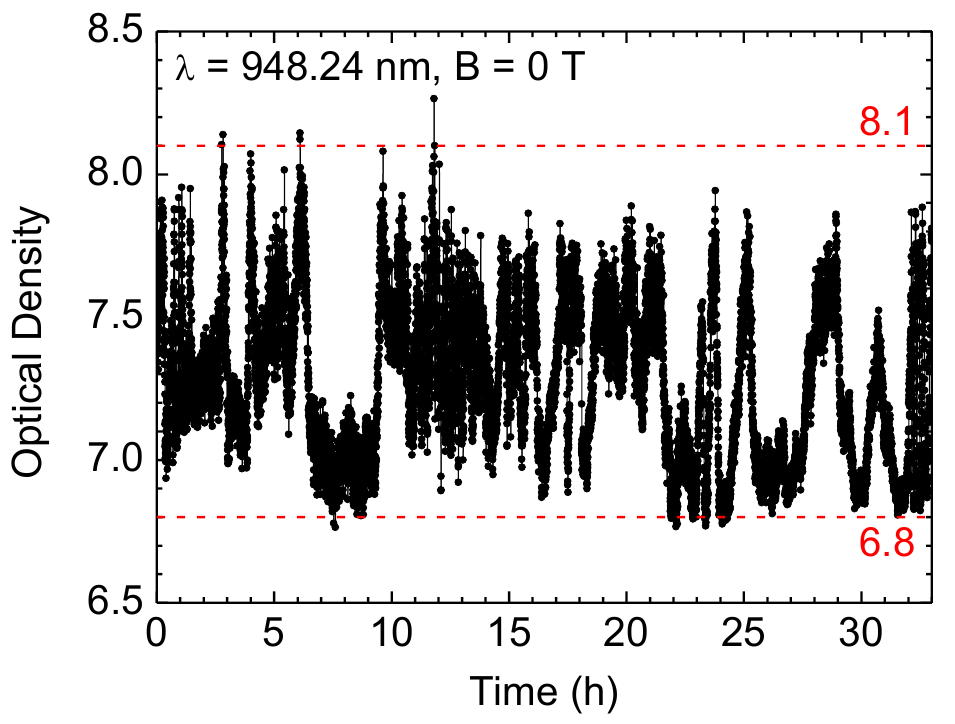}
 \caption{Long-term behaviour. The microscope is aligned to reject the laser reflected at the quantum dot sample (GaAs plus thin metal layer) and the residual counts are recorded by a single photon detector. The optical density (OD), defined as $\rm{OD} = -\log(1/T)$ with transmission $T$, is plotted as a function of time.  The microscope is stable over many hours with an $\rm{OD} > 6.8$.}
 \label{figure4}
 \end{figure}

Despite the high sensitivity to the quarter-wave plate angle (Fig.\ \ref{figure3}), the long-term stability of the microscope is outstanding. It can be operated in a set-and-forget mode: an optical density close to 7, see Fig.\ \ref{figure4}, is achieved over an arbitrary period of time, exceeding typical measurement times by orders of magnitude.   

\section{Resonance fluorescence on a single quantum dot} 
\begin{figure}[t]
\includegraphics[width=\linewidth]{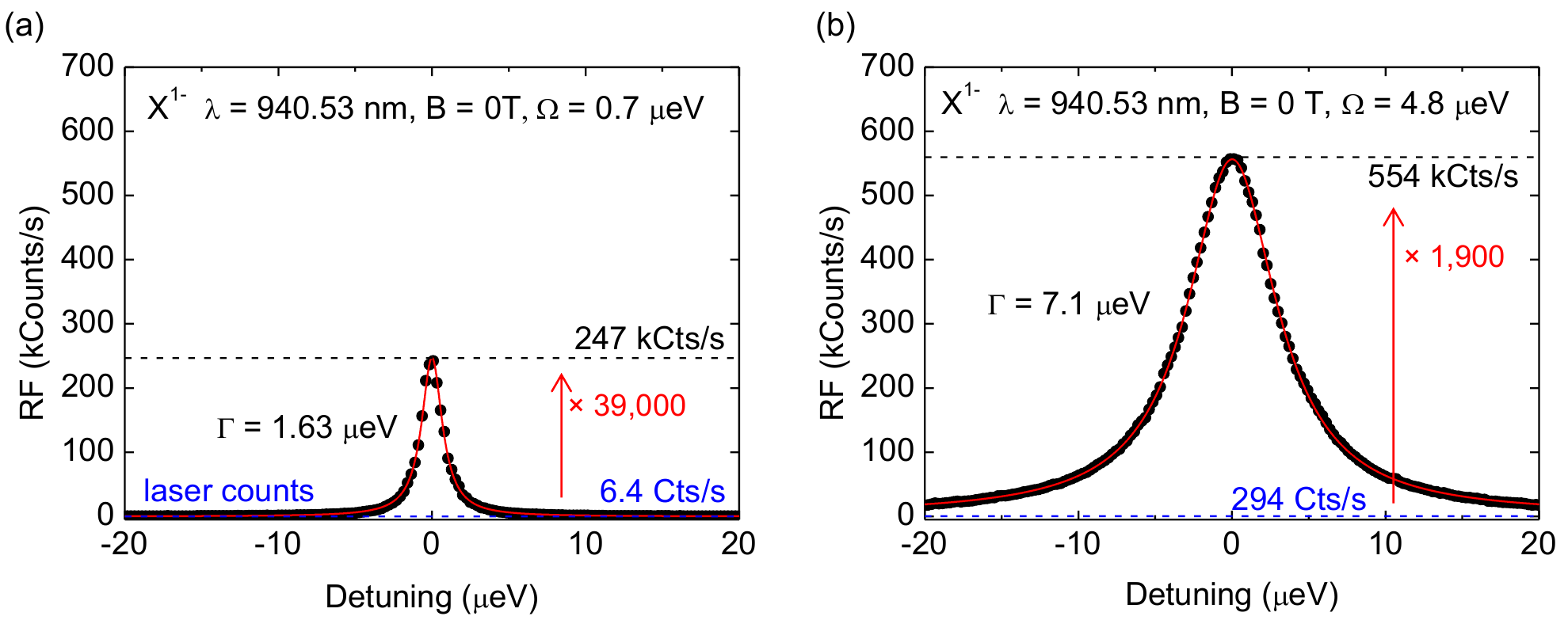}
\caption{Resonance fluorescence on a single InGaAs quantum dot with different optical Rabi couplings. Resonance fluorescence spectra are recorded with a single photon detector at constant laser frequency. Detuning is achieved by sweeping the gate voltage with respect to the laser frequency. (a) Below quantum dot saturation, at an excitation power corresponding to a Rabi energy $\Omega$ of $\unit[0.7]{\mu eV}$, a signal-to-background ratio of $39,000:1$ is achieved. (b) At high pump power, where power broadening dominates the optical linewidth, a signal-to-background ratio $>10^3:1$ is realized. Solid red lines show Lorentzian fits to the data (black points), blue dashed lines indicate the background.}
\label{figure5}
\end{figure}

Once the required high laser suppression is realized, the resonance fluorescence signal-to-background ratio on a single quantum dot is measured. Resonance fluorescence spectra of the single negatively charged exciton X$^{1-}$ recorded at different laser powers and zero magnetic field are shown in Fig.\ \ref{figure5}. The lineshape of the optical resonance is Lorentzian, the full-width at half-maximum (FWHM) is $\unit[1.6]{\mu \rm eV}$ at ``low" power and $\unit[7.1]{\mu \rm eV}$ at ``high" power. The increase in linewidth with power reflects power broadening. Whereas the background, the residual laser signal, increases linearly with laser power, the quantum dot emission saturates and, thus, the signal-to-background ratio is power dependent. At an excitation power below quantum dot saturation the signal-to-background ratio is as high as $39,000:1$ (Fig.\ \ref{figure5} (a)). Above saturation, a ratio $> 10^3:1$ (Fig.\ \ref{figure5} (b)) is achieved. 
 
\begin{figure}[t]
\includegraphics{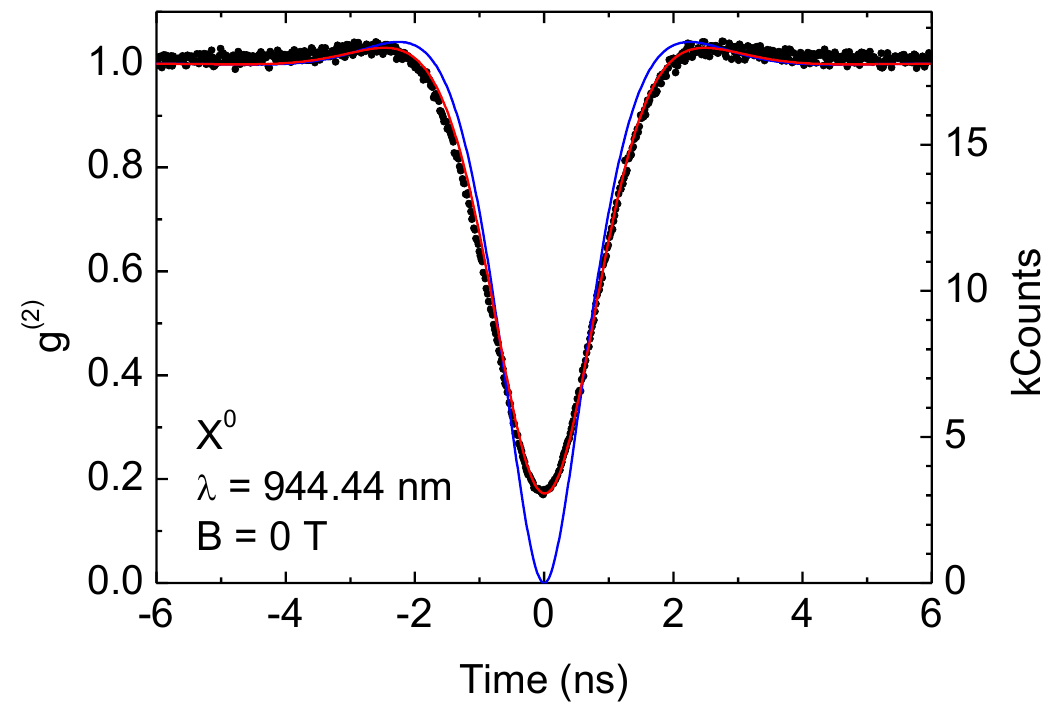}
\caption{$g^{(2)}$ measurement of the resonance fluorescence from the neutral exciton X$^0$ in a single InGaAs quantum dot. A clear dip at zero time delay demonstrates photon anti-bunching. The red curve shows the convolution of the two-level atom result\cite{Loudon2009}, $\rm{g}^{(2)}(t)=1-\left[\cos(\lambda t)+3/\left(2\tau\lambda\right)\sin\left(\lambda t\right)\right] \exp\left[-3 t/(2\tau)\right]$ with $\lambda =  \left(\Omega^2-1/4\tau^2\right)^{1/2}$, Rabi frequency $\Omega$ and radiative lifetime $\tau$, with the response of the detectors (Gaussian with FWHM \unit[0.67]{ns}) and provides a very good description of the data (black points). The blue curve shows the two-level atom response only. A lifetime of $\tau = \unit[(1.0\pm0.1)]{\rm ns}$ and a Rabi frequency $\Omega = \unit[(0.9\pm0.1)]{\mu\rm eV}$ were determined by fitting the data to the convolution. The measurement time was 9 hours with a single channel count rate of $\unit[250]{\rm kHz}$.}
\label{figure6}
\end{figure}

One experiment which requires a high signal-to-background ratio and long integration times (and hence a stable setup) is a $g^{(2)}$ measurement, i.e.\ an intensity correlation experiment. Laser light and a stream of single photons exhibit quite different $g^{(2)}(t=0)$ values, 1 and 0, respectively, such that a leakage of laser light into the single photon stream is very detrimental. The time-dependence of $g^{(2)}$ was measured with a Hanbury Brown-Twiss interferometer (Fig.\ \ref{figure6}). There is a very clear dip at time delay zero, demonstrating anti-bunching in the photon statistics of the neutral exciton X$^0$. Note that even with a single channel count rate of $\unit[250]{\rm kHz}$, an integration time of $\sim 9$ hours was required to achieve a high signal-to-noise ratio in the $g^{(2)}$ measurement: the stability of the dark-field microscope was clearly important. It turns out that the residual value $g^{(2)}(t=0)$ is determined entirely (within the signal:noise) by the jitter in the detectors which is comparable to the radiative decay time. Within error ($\sim 1$\%), the true quantum dot $g^{(2)}(t=0)$ is 0.00. 
 
The resonance fluorescence, presented in Fig.\ \ref{figure5} and Fig.\ \ref{figure6}, was measured on different excitons, the single negatively charged exciton X$^{1-}$ and the neutral X$^0$, respectively. The resonance fluorescence of the X$^{0}$ is linearly polarized ($\pi_x$ or $\pi_y$); the resonance fluorescence of the X$^{1-}$ is unpolarized in the absence of a magnetic field, $B = 0$, circularly polarized ($\sigma^+$ or $\sigma^-$) for $\rm B \ne 0$. The optics of the dark-field microscope define linear s for the excitation and linear p for the detection polarization. Nevertheless, resonance fluorescence of both optically active excitons can be measured independent of the selection rules,  provided that the sample and microscope axes are not aligned. Ideally, the s/p basis is rotated by $\unit[45]{^{\circ}}$ with respect to the $\pi_{x}$/$\pi_{y}$ basis. 

\begin{figure}[t]
\includegraphics{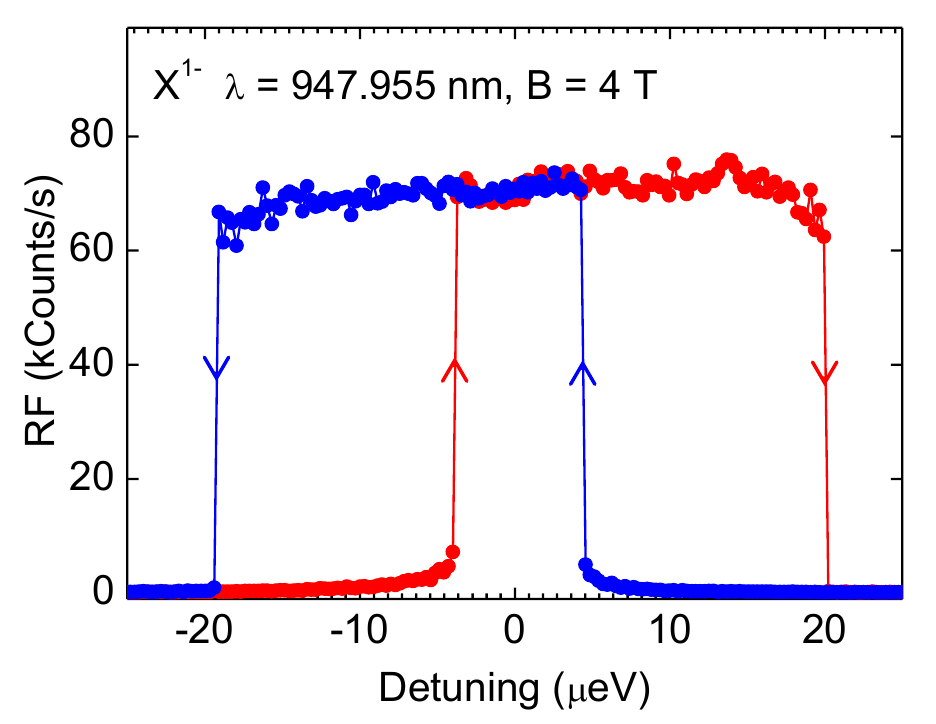}
\caption{Resonance fluorescence spectra of a single InGaAs quantum dot in a magnetic field. The laser suppression at high magnetic field is as good as that achieved at zero magnetic field. At $\rm B=\unit[4]{\rm T}$ the lineshape is a top hat and there is a hysteresis between forward and backward scanning directions. This effect is referred to as dragging\cite{Latta2009,Hogele2012}.}
\label{figure7}
\end{figure}

The dark-field microscope works well across the entire ensemble of quantum dots spanning a bandwidth of about $\unit[60]{\rm nm}$ in wavelength. The dark-field point is so sensitive to the polarization axes that small achromaticities in the polarizers play a role: a change in wavelength $\Delta\lambda$ requires a re-adjustment of the quarter-wave plate and linear polarizer alignments for optimum dark-field performance, typically a few tens of $\unit{\rm m^{\circ}}$ for $\Delta\lambda=\unit[1]{nm}$. Furthermore, resonance fluorescence on a single quantum dot can be recorded not just at $B=0$ but also at high $B$. At high $B$, a high suppression of scattered laser light can be achieved. As for a change in wavelength, for optimum dark-field performance the quarter-wave plate and polarizer alignment have to be re-adjusted as the magnetic field increases. Crucial for the performance at high magnetic field is the linear polarizer angle, differing significantly ($\sim \unit[10]{^{\circ}}$) from the zero field angle probably due to a Faraday effect\cite{Bennett1965} of the objective lens, solid immersion lens or sample. A resonance fluorescence spectrum of an X$^{1-}$ recorded at a magnetic field B of $\unit[4]{T}$ is shown in Fig.\ \ref{figure7}. The lineshape of the optical resonance is clearly non Lorentzian, and there is a hysteresis between forward (red) and backward (blue) detuning. A dynamic nuclear spin polarization locks the quantum resonance to the laser energy as the gate voltage is tuned, an effect referred to as dragging\cite{Latta2009,Hogele2012}. 

As an outlook, we comment that the microscope can be developed further in some simple ways. For instance, given that the quantum dot basis ($\pi_{x}$/$\pi_{y}$) is dot-dependent\cite{Seidl2008}, it may be valuable in the future to include also a way of rotating the microscope basis (s/p) relative to the $\pi_{x}$/$\pi_{y}$ basis, either by inserting an additional wave plate or by rotating the sample. It may also be interesting to develop the capability of operating the microscope not with s/p polarizations but with $\sigma^+$/$\sigma^-$ polarizations. Finally, we note that the quantum efficiency of the resonance fluorescence collection is limited by the high refractive index of the sample: light is refracted to such large angles at the GaAs/vacuum interface that it is collected inefficiently. In this experiment, this situation was rectified to some degree (factor of $\sim 5$ in signal strength) by the solid immersion lens. Despite this low quantum efficiency, the rejection of the scattered laser light in our dark-field microscope is more than sufficient to observe background-free resonance fluorescence from single quantum dots. The next step is therefore to increase the collection efficiency: the dark-field performance is already more than good enough. Candidate structures are resonant micro-cavities, photonic nanowires, or, following the spirit of these experiments, ultra-high index solid immersion lenses. 

% If in two-column mode, this environment will change to single-column format so that long equations can be displayed. 
%\begin{widetext}
%$$\mbox{put long equation here}$$
%\end{widetext}

% Figures should be put into the text as floats. 
% Use the graphics or graphicx packages (distributed with LaTeX2e).
% Put the label that you will use with \ref{} command in the braces of the \label{} command.
%
% \begin{figure}
% \end{figure}

% \begin{table}
% \end{tabular}
% \end{table}

\begin{acknowledgments}
We acknowledge financial support from NCCR QSIT. We thank Paul Dalgarno for helpful discussions; Sascha Martin and Michael Steinacher for technical support. D.B. acknowledges support from the Marie-Curie Society, Project \#275840. A.L., D.R. and A.D.W. acknowledge gratefully support from DFG SPP1285 and BMBF QuaHLRep 01BQ1035. 
\end{acknowledgments}

\newpage\clearpage


\begin{thebibliography}{31}%
\makeatletter
\providecommand \@ifxundefined [1]{%
 \@ifx{#1\undefined}
}%
\providecommand \@ifnum [1]{%
 \ifnum #1\expandafter \@firstoftwo
 \else \expandafter \@secondoftwo
 \fi
}%
\providecommand \@ifx [1]{%
 \ifx #1\expandafter \@firstoftwo
 \else \expandafter \@secondoftwo
 \fi
}%
\providecommand \natexlab [1]{#1}%
\providecommand \enquote  [1]{``#1''}%
\providecommand \bibnamefont  [1]{#1}%
\providecommand \bibfnamefont [1]{#1}%
\providecommand \citenamefont [1]{#1}%
\providecommand \href@noop [0]{\@secondoftwo}%
\providecommand \href [0]{\begingroup \@sanitize@url \@href}%
\providecommand \@href[1]{\@@startlink{#1}\@@href}%
\providecommand \@@href[1]{\endgroup#1\@@endlink}%
\providecommand \@sanitize@url [0]{\catcode `\\12\catcode `\$12\catcode
  `\&12\catcode `\#12\catcode `\^12\catcode `\_12\catcode `\%12\relax}%
\providecommand \@@startlink[1]{}%
\providecommand \@@endlink[0]{}%
\providecommand \url  [0]{\begingroup\@sanitize@url \@url }%
\providecommand \@url [1]{\endgroup\@href {#1}{\urlprefix }}%
\providecommand \urlprefix  [0]{URL }%
\providecommand \Eprint [0]{\href }%
\providecommand \doibase [0]{http://dx.doi.org/}%
\providecommand \selectlanguage [0]{\@gobble}%
\providecommand \bibinfo  [0]{\@secondoftwo}%
\providecommand \bibfield  [0]{\@secondoftwo}%
\providecommand \translation [1]{[#1]}%
\providecommand \BibitemOpen [0]{}%
\providecommand \bibitemStop [0]{}%
\providecommand \bibitemNoStop [0]{.\EOS\space}%
\providecommand \EOS [0]{\spacefactor3000\relax}%
\providecommand \BibitemShut  [1]{\csname bibitem#1\endcsname}%
\let\auto@bib@innerbib\@empty
%</preamble>
\bibitem [{\citenamefont {Shields}(2007)}]{Shields2007}%
  \BibitemOpen
  \bibfield  {author} {\bibinfo {author} {\bibfnamefont {A.~J.}\ \bibnamefont
  {Shields}},\ }\href@noop {} {\bibfield  {journal} {\bibinfo  {journal}
  {Nature Photonics}\ }\textbf {\bibinfo {volume} {1}},\ \bibinfo {pages} {215}
  (\bibinfo {year} {2007})}\BibitemShut {NoStop}%
\bibitem [{\citenamefont {Warburton}(2013)}]{Warburton2013}%
  \BibitemOpen
  \bibfield  {author} {\bibinfo {author} {\bibfnamefont {R.~J.}\ \bibnamefont
  {Warburton}},\ }\href@noop {} {\bibfield  {journal} {\bibinfo  {journal}
  {Nature Materials}\ } (\bibinfo {year} {2013})},\ \bibinfo {note} {to be
  published}\BibitemShut {NoStop}%
\bibitem [{\citenamefont {Kuhlmann}\ \emph {et~al.}(2013)\citenamefont
  {Kuhlmann}, \citenamefont {Houel}, \citenamefont {Ludwig}, \citenamefont
  {Greuter}, \citenamefont {Reuter}, \citenamefont {Wieck}, \citenamefont
  {Poggio},\ and\ \citenamefont {Warburton}}]{Kuhlmann2013}%
  \BibitemOpen
  \bibfield  {author} {\bibinfo {author} {\bibfnamefont {A.~V.}\ \bibnamefont
  {Kuhlmann}}, \bibinfo {author} {\bibfnamefont {J.}~\bibnamefont {Houel}},
  \bibinfo {author} {\bibfnamefont {A.}~\bibnamefont {Ludwig}}, \bibinfo
  {author} {\bibfnamefont {L.}~\bibnamefont {Greuter}}, \bibinfo {author}
  {\bibfnamefont {D.}~\bibnamefont {Reuter}}, \bibinfo {author} {\bibfnamefont
  {A.~D.}\ \bibnamefont {Wieck}}, \bibinfo {author} {\bibfnamefont
  {M.}~\bibnamefont {Poggio}}, \ and\ \bibinfo {author} {\bibfnamefont {R.~J.}\
  \bibnamefont {Warburton}},\ }\href@noop {} {\bibfield  {journal} {\bibinfo
  {journal} {arXiv:1301.6381}\ } (\bibinfo {year} {2013})}\BibitemShut
  {NoStop}%
\bibitem [{\citenamefont {Atat\"ure}\ \emph {et~al.}(2006)\citenamefont
  {Atat\"ure}, \citenamefont {Dreiser}, \citenamefont {Badolato}, \citenamefont
  {H\"ogele}, \citenamefont {Karrai},\ and\ \citenamefont
  {Imamoglu}}]{Atature2006}%
  \BibitemOpen
  \bibfield  {author} {\bibinfo {author} {\bibfnamefont {M.}~\bibnamefont
  {Atat\"ure}}, \bibinfo {author} {\bibfnamefont {J.}~\bibnamefont {Dreiser}},
  \bibinfo {author} {\bibfnamefont {A.}~\bibnamefont {Badolato}}, \bibinfo
  {author} {\bibfnamefont {A.}~\bibnamefont {H\"ogele}}, \bibinfo {author}
  {\bibfnamefont {K.}~\bibnamefont {Karrai}}, \ and\ \bibinfo {author}
  {\bibfnamefont {A.}~\bibnamefont {Imamoglu}},\ }\href {\doibase
  10.1126/science.1126074} {\bibfield  {journal} {\bibinfo  {journal}
  {Science}\ }\textbf {\bibinfo {volume} {312}},\ \bibinfo {pages} {551}
  (\bibinfo {year} {2006})}\BibitemShut {NoStop}%
\bibitem [{\citenamefont {Gerardot}\ \emph {et~al.}(2008)\citenamefont
  {Gerardot}, \citenamefont {Brunner}, \citenamefont {Dalgarno}, \citenamefont
  {Ohberg}, \citenamefont {Seidl}, \citenamefont {Kroner}, \citenamefont
  {Karrai}, \citenamefont {Stoltz}, \citenamefont {Petroff},\ and\
  \citenamefont {Warburton}}]{Gerardot2008}%
  \BibitemOpen
  \bibfield  {author} {\bibinfo {author} {\bibfnamefont {B.~D.}\ \bibnamefont
  {Gerardot}}, \bibinfo {author} {\bibfnamefont {D.}~\bibnamefont {Brunner}},
  \bibinfo {author} {\bibfnamefont {P.~A.}\ \bibnamefont {Dalgarno}}, \bibinfo
  {author} {\bibfnamefont {P.}~\bibnamefont {Ohberg}}, \bibinfo {author}
  {\bibfnamefont {S.}~\bibnamefont {Seidl}}, \bibinfo {author} {\bibfnamefont
  {M.}~\bibnamefont {Kroner}}, \bibinfo {author} {\bibfnamefont
  {K.}~\bibnamefont {Karrai}}, \bibinfo {author} {\bibfnamefont {N.~G.}\
  \bibnamefont {Stoltz}}, \bibinfo {author} {\bibfnamefont {P.~M.}\
  \bibnamefont {Petroff}}, \ and\ \bibinfo {author} {\bibfnamefont {R.~J.}\
  \bibnamefont {Warburton}},\ }\href {\doibase 10.1038/nature06472} {\bibfield
  {journal} {\bibinfo  {journal} {Nature}\ }\textbf {\bibinfo {volume} {451}},\
  \bibinfo {pages} {441} (\bibinfo {year} {2008})}\BibitemShut {NoStop}%
\bibitem [{\citenamefont {Press}\ \emph {et~al.}(2008)\citenamefont {Press},
  \citenamefont {Ladd}, \citenamefont {Zhang},\ and\ \citenamefont
  {Yamamoto}}]{Press2008}%
  \BibitemOpen
  \bibfield  {author} {\bibinfo {author} {\bibfnamefont {D.}~\bibnamefont
  {Press}}, \bibinfo {author} {\bibfnamefont {T.~D.}\ \bibnamefont {Ladd}},
  \bibinfo {author} {\bibfnamefont {B.}~\bibnamefont {Zhang}}, \ and\ \bibinfo
  {author} {\bibfnamefont {Y.}~\bibnamefont {Yamamoto}},\ }\href {\doibase
  10.1038/nature07530} {\bibfield  {journal} {\bibinfo  {journal} {Nature}\
  }\textbf {\bibinfo {volume} {456}},\ \bibinfo {pages} {218} (\bibinfo {year}
  {2008})}\BibitemShut {NoStop}%
\bibitem [{\citenamefont {Vamivakas}\ and\ \citenamefont
  {Atat{\"u}re}(2010)}]{Vamivakas2010}%
  \BibitemOpen
  \bibfield  {author} {\bibinfo {author} {\bibfnamefont {A.~N.}\ \bibnamefont
  {Vamivakas}}\ and\ \bibinfo {author} {\bibfnamefont {M.}~\bibnamefont
  {Atat{\"u}re}},\ }\href {\doibase 10.1080/00107510903298198} {\bibfield
  {journal} {\bibinfo  {journal} {Contemporary Physics}\ }\textbf {\bibinfo
  {volume} {51}},\ \bibinfo {pages} {17} (\bibinfo {year} {2010})}\BibitemShut
  {NoStop}%
\bibitem [{\citenamefont {H\"ogele}\ \emph {et~al.}(2004)\citenamefont
  {H\"ogele}, \citenamefont {Seidl}, \citenamefont {Kroner}, \citenamefont
  {Karrai}, \citenamefont {Warburton}, \citenamefont {Gerardot},\ and\
  \citenamefont {Petroff}}]{Hogele2004}%
  \BibitemOpen
  \bibfield  {author} {\bibinfo {author} {\bibfnamefont {A.}~\bibnamefont
  {H\"ogele}}, \bibinfo {author} {\bibfnamefont {S.}~\bibnamefont {Seidl}},
  \bibinfo {author} {\bibfnamefont {M.}~\bibnamefont {Kroner}}, \bibinfo
  {author} {\bibfnamefont {K.}~\bibnamefont {Karrai}}, \bibinfo {author}
  {\bibfnamefont {R.~J.}\ \bibnamefont {Warburton}}, \bibinfo {author}
  {\bibfnamefont {B.~D.}\ \bibnamefont {Gerardot}}, \ and\ \bibinfo {author}
  {\bibfnamefont {P.~M.}\ \bibnamefont {Petroff}},\ }\href {\doibase
  10.1103/PhysRevLett.93.217401} {\bibfield  {journal} {\bibinfo  {journal}
  {Phys. Rev. Lett.}\ }\textbf {\bibinfo {volume} {93}},\ \bibinfo {pages}
  {217401} (\bibinfo {year} {2004})}\BibitemShut {NoStop}%
\bibitem [{\citenamefont {Al\'{e}n}\ \emph {et~al.}(2003)\citenamefont
  {Al\'{e}n}, \citenamefont {Bickel}, \citenamefont {Karrai}, \citenamefont
  {Warburton},\ and\ \citenamefont {Petroff}}]{Al'en2003}%
  \BibitemOpen
  \bibfield  {author} {\bibinfo {author} {\bibfnamefont {B.}~\bibnamefont
  {Al\'{e}n}}, \bibinfo {author} {\bibfnamefont {F.}~\bibnamefont {Bickel}},
  \bibinfo {author} {\bibfnamefont {K.}~\bibnamefont {Karrai}}, \bibinfo
  {author} {\bibfnamefont {R.~J.}\ \bibnamefont {Warburton}}, \ and\ \bibinfo
  {author} {\bibfnamefont {P.~M.}\ \bibnamefont {Petroff}},\ }\href {\doibase
  10.1063/1.1609243} {\bibfield  {journal} {\bibinfo  {journal} {App. Phys.
  Lett.}\ }\textbf {\bibinfo {volume} {83}},\ \bibinfo {pages} {2235} (\bibinfo
  {year} {2003})}\BibitemShut {NoStop}%
\bibitem [{\citenamefont {Karrai}\ and\ \citenamefont
  {Warburton}(2003)}]{Karrai2003}%
  \BibitemOpen
  \bibfield  {author} {\bibinfo {author} {\bibfnamefont {K.}~\bibnamefont
  {Karrai}}\ and\ \bibinfo {author} {\bibfnamefont {R.~J.}\ \bibnamefont
  {Warburton}},\ }\href {\doibase DOI: 10.1016/j.spmi.2004.02.007} {\bibfield
  {journal} {\bibinfo  {journal} {Superlattices and Microstructures}\ }\textbf
  {\bibinfo {volume} {33}},\ \bibinfo {pages} {311 } (\bibinfo {year}
  {2003})}\BibitemShut {NoStop}%
\bibitem [{\citenamefont {Kimble}(2008)}]{Kimble2008}%
  \BibitemOpen
  \bibfield  {author} {\bibinfo {author} {\bibfnamefont {H.~J.}\ \bibnamefont
  {Kimble}},\ }\href@noop {} {\bibfield  {journal} {\bibinfo  {journal}
  {Nature}\ }\textbf {\bibinfo {volume} {453}},\ \bibinfo {pages} {1023}
  (\bibinfo {year} {2008})}\BibitemShut {NoStop}%
\bibitem [{\citenamefont {Muller}\ \emph {et~al.}(2007)\citenamefont {Muller},
  \citenamefont {Flagg}, \citenamefont {Bianucci}, \citenamefont {Wang},
  \citenamefont {Deppe}, \citenamefont {Ma}, \citenamefont {Zhang},
  \citenamefont {Salamo}, \citenamefont {Xiao},\ and\ \citenamefont
  {Shih}}]{Muller2007}%
  \BibitemOpen
  \bibfield  {author} {\bibinfo {author} {\bibfnamefont {A.}~\bibnamefont
  {Muller}}, \bibinfo {author} {\bibfnamefont {E.~B.}\ \bibnamefont {Flagg}},
  \bibinfo {author} {\bibfnamefont {P.}~\bibnamefont {Bianucci}}, \bibinfo
  {author} {\bibfnamefont {X.~Y.}\ \bibnamefont {Wang}}, \bibinfo {author}
  {\bibfnamefont {D.~G.}\ \bibnamefont {Deppe}}, \bibinfo {author}
  {\bibfnamefont {W.}~\bibnamefont {Ma}}, \bibinfo {author} {\bibfnamefont
  {J.}~\bibnamefont {Zhang}}, \bibinfo {author} {\bibfnamefont {G.~J.}\
  \bibnamefont {Salamo}}, \bibinfo {author} {\bibfnamefont {M.}~\bibnamefont
  {Xiao}}, \ and\ \bibinfo {author} {\bibfnamefont {C.~K.}\ \bibnamefont
  {Shih}},\ }\href {\doibase 10.1103/PhysRevLett.99.187402} {\bibfield
  {journal} {\bibinfo  {journal} {Phys. Rev. Lett.}\ }\textbf {\bibinfo
  {volume} {99}},\ \bibinfo {pages} {187402} (\bibinfo {year}
  {2007})}\BibitemShut {NoStop}%
\bibitem [{\citenamefont {Ates}\ \emph {et~al.}(2009)\citenamefont {Ates},
  \citenamefont {Ulrich}, \citenamefont {Reitzenstein}, \citenamefont
  {L\"offler}, \citenamefont {Forchel},\ and\ \citenamefont
  {Michler}}]{Ates2009}%
  \BibitemOpen
  \bibfield  {author} {\bibinfo {author} {\bibfnamefont {S.}~\bibnamefont
  {Ates}}, \bibinfo {author} {\bibfnamefont {S.~M.}\ \bibnamefont {Ulrich}},
  \bibinfo {author} {\bibfnamefont {S.}~\bibnamefont {Reitzenstein}}, \bibinfo
  {author} {\bibfnamefont {A.}~\bibnamefont {L\"offler}}, \bibinfo {author}
  {\bibfnamefont {A.}~\bibnamefont {Forchel}}, \ and\ \bibinfo {author}
  {\bibfnamefont {P.}~\bibnamefont {Michler}},\ }\href {\doibase
  10.1103/PhysRevLett.103.167402} {\bibfield  {journal} {\bibinfo  {journal}
  {Phys. Rev. Lett.}\ }\textbf {\bibinfo {volume} {103}},\ \bibinfo {pages}
  {167402} (\bibinfo {year} {2009})}\BibitemShut {NoStop}%
\bibitem [{\citenamefont {Nguyen}\ \emph {et~al.}(2011)\citenamefont {Nguyen},
  \citenamefont {Sallen}, \citenamefont {Voisin}, \citenamefont {Roussignol},
  \citenamefont {Diederichs},\ and\ \citenamefont {Cassabois}}]{Nguyen2011}%
  \BibitemOpen
  \bibfield  {author} {\bibinfo {author} {\bibfnamefont {H.~S.}\ \bibnamefont
  {Nguyen}}, \bibinfo {author} {\bibfnamefont {G.}~\bibnamefont {Sallen}},
  \bibinfo {author} {\bibfnamefont {C.}~\bibnamefont {Voisin}}, \bibinfo
  {author} {\bibfnamefont {P.}~\bibnamefont {Roussignol}}, \bibinfo {author}
  {\bibfnamefont {C.}~\bibnamefont {Diederichs}}, \ and\ \bibinfo {author}
  {\bibfnamefont {G.}~\bibnamefont {Cassabois}},\ }\href {\doibase
  10.1063/1.3672034} {\bibfield  {journal} {\bibinfo  {journal} {App. Phys.
  Lett.}\ }\textbf {\bibinfo {volume} {99}},\ \bibinfo {eid} {261904} (\bibinfo
  {year} {2011})}\BibitemShut {NoStop}%
\bibitem [{\citenamefont {Ulhaq}\ \emph {et~al.}(2012)\citenamefont {Ulhaq},
  \citenamefont {Weiler}, \citenamefont {Ulrich}, \citenamefont {Roszbach},
  \citenamefont {Jetter},\ and\ \citenamefont {Michler}}]{Ulhaq2012}%
  \BibitemOpen
  \bibfield  {author} {\bibinfo {author} {\bibfnamefont {A.}~\bibnamefont
  {Ulhaq}}, \bibinfo {author} {\bibfnamefont {S.}~\bibnamefont {Weiler}},
  \bibinfo {author} {\bibfnamefont {S.~M.}\ \bibnamefont {Ulrich}}, \bibinfo
  {author} {\bibfnamefont {R.}~\bibnamefont {Roszbach}}, \bibinfo {author}
  {\bibfnamefont {M.}~\bibnamefont {Jetter}}, \ and\ \bibinfo {author}
  {\bibfnamefont {P.}~\bibnamefont {Michler}},\ }\href@noop {} {\bibfield
  {journal} {\bibinfo  {journal} {Nature Photonics}\ }\textbf {\bibinfo
  {volume} {6}},\ \bibinfo {pages} {238} (\bibinfo {year} {2012})}\BibitemShut
  {NoStop}%
\bibitem [{\citenamefont {Vamivakas}\ \emph {et~al.}(2009)\citenamefont
  {Vamivakas}, \citenamefont {Zhao}, \citenamefont {Lu},\ and\ \citenamefont
  {Atat\"ure}}]{Vamivakas2009}%
  \BibitemOpen
  \bibfield  {author} {\bibinfo {author} {\bibfnamefont {A.~N.}\ \bibnamefont
  {Vamivakas}}, \bibinfo {author} {\bibfnamefont {Y.}~\bibnamefont {Zhao}},
  \bibinfo {author} {\bibfnamefont {C.~Y.}\ \bibnamefont {Lu}}, \ and\ \bibinfo
  {author} {\bibfnamefont {M.}~\bibnamefont {Atat\"ure}},\ }\href@noop {}
  {\bibfield  {journal} {\bibinfo  {journal} {Nature Physics}\ }\textbf
  {\bibinfo {volume} {5}},\ \bibinfo {pages} {198 } (\bibinfo {year}
  {2009})}\BibitemShut {NoStop}%
\bibitem [{\citenamefont {Vamivakas}\ \emph {et~al.}(2010)\citenamefont
  {Vamivakas}, \citenamefont {Lu}, \citenamefont {Matthiesen}, \citenamefont
  {Zhao}, \citenamefont {F{\"a}lt}, \citenamefont {Badolato},\ and\
  \citenamefont {Atat{\"u}re}}]{Vamivakas2010a}%
  \BibitemOpen
  \bibfield  {author} {\bibinfo {author} {\bibfnamefont {A.~N.}\ \bibnamefont
  {Vamivakas}}, \bibinfo {author} {\bibfnamefont {C.~Y.}\ \bibnamefont {Lu}},
  \bibinfo {author} {\bibfnamefont {C.}~\bibnamefont {Matthiesen}}, \bibinfo
  {author} {\bibfnamefont {Y.}~\bibnamefont {Zhao}}, \bibinfo {author}
  {\bibfnamefont {S.}~\bibnamefont {F{\"a}lt}}, \bibinfo {author}
  {\bibfnamefont {A.}~\bibnamefont {Badolato}}, \ and\ \bibinfo {author}
  {\bibfnamefont {M.}~\bibnamefont {Atat{\"u}re}},\ }\href@noop {} {\bibfield
  {journal} {\bibinfo  {journal} {Nature}\ }\textbf {\bibinfo {volume} {467}},\
  \bibinfo {pages} {297} (\bibinfo {year} {2010})}\BibitemShut {NoStop}%
\bibitem [{\citenamefont {Yilmaz}, \citenamefont {Fallahi},\ and\ \citenamefont
  {Imamoglu}(2010)}]{Yilmaz2010}%
  \BibitemOpen
  \bibfield  {author} {\bibinfo {author} {\bibfnamefont {S.~T.}\ \bibnamefont
  {Yilmaz}}, \bibinfo {author} {\bibfnamefont {P.}~\bibnamefont {Fallahi}}, \
  and\ \bibinfo {author} {\bibfnamefont {A.}~\bibnamefont {Imamoglu}},\ }\href
  {\doibase 10.1103/PhysRevLett.105.033601} {\bibfield  {journal} {\bibinfo
  {journal} {Phys. Rev. Lett.}\ }\textbf {\bibinfo {volume} {105}},\ \bibinfo
  {pages} {033601} (\bibinfo {year} {2010})}\BibitemShut {NoStop}%
\bibitem [{\citenamefont {Matthiesen}, \citenamefont {Vamivakas},\ and\
  \citenamefont {Atat\"ure}(2012)}]{Matthiesen2012}%
  \BibitemOpen
  \bibfield  {author} {\bibinfo {author} {\bibfnamefont {C.}~\bibnamefont
  {Matthiesen}}, \bibinfo {author} {\bibfnamefont {A.~N.}\ \bibnamefont
  {Vamivakas}}, \ and\ \bibinfo {author} {\bibfnamefont {M.}~\bibnamefont
  {Atat\"ure}},\ }\href {\doibase 10.1103/PhysRevLett.108.093602} {\bibfield
  {journal} {\bibinfo  {journal} {Phys. Rev. Lett.}\ }\textbf {\bibinfo
  {volume} {108}},\ \bibinfo {pages} {093602} (\bibinfo {year}
  {2012})}\BibitemShut {NoStop}%
\bibitem [{\citenamefont {Houel}\ \emph {et~al.}(2012)\citenamefont {Houel},
  \citenamefont {Kuhlmann}, \citenamefont {Greuter}, \citenamefont {Xue},
  \citenamefont {Poggio}, \citenamefont {Gerardot}, \citenamefont {Dalgarno},
  \citenamefont {Badolato}, \citenamefont {Petroff}, \citenamefont {Ludwig},
  \citenamefont {Reuter}, \citenamefont {Wieck},\ and\ \citenamefont
  {Warburton}}]{Houel2012}%
  \BibitemOpen
  \bibfield  {author} {\bibinfo {author} {\bibfnamefont {J.}~\bibnamefont
  {Houel}}, \bibinfo {author} {\bibfnamefont {A.~V.}\ \bibnamefont {Kuhlmann}},
  \bibinfo {author} {\bibfnamefont {L.}~\bibnamefont {Greuter}}, \bibinfo
  {author} {\bibfnamefont {F.}~\bibnamefont {Xue}}, \bibinfo {author}
  {\bibfnamefont {M.}~\bibnamefont {Poggio}}, \bibinfo {author} {\bibfnamefont
  {B.~D.}\ \bibnamefont {Gerardot}}, \bibinfo {author} {\bibfnamefont {P.~A.}\
  \bibnamefont {Dalgarno}}, \bibinfo {author} {\bibfnamefont {A.}~\bibnamefont
  {Badolato}}, \bibinfo {author} {\bibfnamefont {P.~M.}\ \bibnamefont
  {Petroff}}, \bibinfo {author} {\bibfnamefont {A.}~\bibnamefont {Ludwig}},
  \bibinfo {author} {\bibfnamefont {D.}~\bibnamefont {Reuter}}, \bibinfo
  {author} {\bibfnamefont {A.~D.}\ \bibnamefont {Wieck}}, \ and\ \bibinfo
  {author} {\bibfnamefont {R.~J.}\ \bibnamefont {Warburton}},\ }\href {\doibase
  10.1103/PhysRevLett.108.107401} {\bibfield  {journal} {\bibinfo  {journal}
  {Phys. Rev. Lett.}\ }\textbf {\bibinfo {volume} {108}},\ \bibinfo {pages}
  {107401} (\bibinfo {year} {2012})}\BibitemShut {NoStop}%
\bibitem [{\citenamefont {Gao}\ \emph {et~al.}(2012)\citenamefont {Gao},
  \citenamefont {Fallahi}, \citenamefont {Togan}, \citenamefont
  {Miguel-Sanchez},\ and\ \citenamefont {Imamoglu}}]{Gao2012}%
  \BibitemOpen
  \bibfield  {author} {\bibinfo {author} {\bibfnamefont {W.~B.}\ \bibnamefont
  {Gao}}, \bibinfo {author} {\bibfnamefont {P.}~\bibnamefont {Fallahi}},
  \bibinfo {author} {\bibfnamefont {E.}~\bibnamefont {Togan}}, \bibinfo
  {author} {\bibfnamefont {J.}~\bibnamefont {Miguel-Sanchez}}, \ and\ \bibinfo
  {author} {\bibfnamefont {A.}~\bibnamefont {Imamoglu}},\ }\href@noop {}
  {\bibfield  {journal} {\bibinfo  {journal} {Nature}\ }\textbf {\bibinfo
  {volume} {491}},\ \bibinfo {pages} {426} (\bibinfo {year}
  {2012})}\BibitemShut {NoStop}%
\bibitem [{\citenamefont {Schuda}, \citenamefont {Jr},\ and\ \citenamefont
  {Hercher}(1974)}]{Schuda1974}%
  \BibitemOpen
  \bibfield  {author} {\bibinfo {author} {\bibfnamefont {F.}~\bibnamefont
  {Schuda}}, \bibinfo {author} {\bibfnamefont {C.~R.~S.}\ \bibnamefont {Jr}}, \
  and\ \bibinfo {author} {\bibfnamefont {M.}~\bibnamefont {Hercher}},\ }\href
  {http://stacks.iop.org/0022-3700/7/i=7/a=002} {\bibfield  {journal} {\bibinfo
   {journal} {Journal of Physics B: Atomic and Molecular Physics}\ }\textbf
  {\bibinfo {volume} {7}},\ \bibinfo {pages} {L198} (\bibinfo {year}
  {1974})}\BibitemShut {NoStop}%
\bibitem [{\citenamefont {Wu}, \citenamefont {Grove},\ and\ \citenamefont
  {Ezekiel}(1975)}]{Wu1975}%
  \BibitemOpen
  \bibfield  {author} {\bibinfo {author} {\bibfnamefont {F.~Y.}\ \bibnamefont
  {Wu}}, \bibinfo {author} {\bibfnamefont {R.~E.}\ \bibnamefont {Grove}}, \
  and\ \bibinfo {author} {\bibfnamefont {S.}~\bibnamefont {Ezekiel}},\ }\href
  {\doibase 10.1103/PhysRevLett.35.1426} {\bibfield  {journal} {\bibinfo
  {journal} {Phys. Rev. Lett.}\ }\textbf {\bibinfo {volume} {35}},\ \bibinfo
  {pages} {1426} (\bibinfo {year} {1975})}\BibitemShut {NoStop}%
\bibitem [{\citenamefont {Drexler}\ \emph {et~al.}(1994)\citenamefont
  {Drexler}, \citenamefont {Leonard}, \citenamefont {Hansen}, \citenamefont
  {Kotthaus},\ and\ \citenamefont {Petroff}}]{Drexler1994}%
  \BibitemOpen
  \bibfield  {author} {\bibinfo {author} {\bibfnamefont {H.}~\bibnamefont
  {Drexler}}, \bibinfo {author} {\bibfnamefont {D.}~\bibnamefont {Leonard}},
  \bibinfo {author} {\bibfnamefont {W.}~\bibnamefont {Hansen}}, \bibinfo
  {author} {\bibfnamefont {J.~P.}\ \bibnamefont {Kotthaus}}, \ and\ \bibinfo
  {author} {\bibfnamefont {P.~M.}\ \bibnamefont {Petroff}},\ }\href {\doibase
  10.1103/PhysRevLett.73.2252} {\bibfield  {journal} {\bibinfo  {journal}
  {Phys. Rev. Lett.}\ }\textbf {\bibinfo {volume} {73}},\ \bibinfo {pages}
  {2252} (\bibinfo {year} {1994})}\BibitemShut {NoStop}%
\bibitem [{\citenamefont {Warburton}\ \emph {et~al.}(2000)\citenamefont
  {Warburton}, \citenamefont {Sch\"{a}flein}, \citenamefont {Haft},
  \citenamefont {Bickel}, \citenamefont {Lorke}, \citenamefont {Karrai},
  \citenamefont {Garcia}, \citenamefont {Schoenfeld},\ and\ \citenamefont
  {Petroff}}]{Warburton2000}%
  \BibitemOpen
  \bibfield  {author} {\bibinfo {author} {\bibfnamefont {R.~J.}\ \bibnamefont
  {Warburton}}, \bibinfo {author} {\bibfnamefont {C.}~\bibnamefont
  {Sch\"{a}flein}}, \bibinfo {author} {\bibfnamefont {D.}~\bibnamefont {Haft}},
  \bibinfo {author} {\bibfnamefont {F.}~\bibnamefont {Bickel}}, \bibinfo
  {author} {\bibfnamefont {A.}~\bibnamefont {Lorke}}, \bibinfo {author}
  {\bibfnamefont {K.}~\bibnamefont {Karrai}}, \bibinfo {author} {\bibfnamefont
  {J.}~\bibnamefont {Garcia}}, \bibinfo {author} {\bibfnamefont
  {W.}~\bibnamefont {Schoenfeld}}, \ and\ \bibinfo {author} {\bibfnamefont
  {P.}~\bibnamefont {Petroff}},\ }\href@noop {} {\bibfield  {journal} {\bibinfo
   {journal} {Nature}\ }\textbf {\bibinfo {volume} {405}},\ \bibinfo {pages}
  {926 } (\bibinfo {year} {2000})}\BibitemShut {NoStop}%
\bibitem [{\citenamefont {Gerardot}\ \emph {et~al.}(2007)\citenamefont
  {Gerardot}, \citenamefont {Seidl}, \citenamefont {Dalgarno}, \citenamefont
  {Warburton}, \citenamefont {Kroner}, \citenamefont {Karrai}, \citenamefont
  {Badolato},\ and\ \citenamefont {Petroff}}]{Gerardot2007}%
  \BibitemOpen
  \bibfield  {author} {\bibinfo {author} {\bibfnamefont {B.~D.}\ \bibnamefont
  {Gerardot}}, \bibinfo {author} {\bibfnamefont {S.}~\bibnamefont {Seidl}},
  \bibinfo {author} {\bibfnamefont {P.~A.}\ \bibnamefont {Dalgarno}}, \bibinfo
  {author} {\bibfnamefont {R.~J.}\ \bibnamefont {Warburton}}, \bibinfo {author}
  {\bibfnamefont {M.}~\bibnamefont {Kroner}}, \bibinfo {author} {\bibfnamefont
  {K.}~\bibnamefont {Karrai}}, \bibinfo {author} {\bibfnamefont
  {A.}~\bibnamefont {Badolato}}, \ and\ \bibinfo {author} {\bibfnamefont
  {P.~M.}\ \bibnamefont {Petroff}},\ }\href {\doibase 10.1063/1.2743750}
  {\bibfield  {journal} {\bibinfo  {journal} {App. Phys. Lett.}\ }\textbf
  {\bibinfo {volume} {90}},\ \bibinfo {eid} {221106} (\bibinfo {year}
  {2007})}\BibitemShut {NoStop}%
\bibitem [{\citenamefont {Loudon}(2009)}]{Loudon2009}%
  \BibitemOpen
  \bibfield  {author} {\bibinfo {author} {\bibfnamefont {R.}~\bibnamefont
  {Loudon}},\ }\href@noop {} {\emph {\bibinfo {title} {The Quantum Theory of
  light}}}\ (\bibinfo  {publisher} {Oxford Science Publications},\ \bibinfo
  {year} {2009})\BibitemShut {NoStop}%
\bibitem [{\citenamefont {Bennett}\ and\ \citenamefont
  {Stern}(1965)}]{Bennett1965}%
  \BibitemOpen
  \bibfield  {author} {\bibinfo {author} {\bibfnamefont {H.~S.}\ \bibnamefont
  {Bennett}}\ and\ \bibinfo {author} {\bibfnamefont {E.~A.}\ \bibnamefont
  {Stern}},\ }\href {\doibase 10.1103/PhysRev.137.A448} {\bibfield  {journal}
  {\bibinfo  {journal} {Phys. Rev.}\ }\textbf {\bibinfo {volume} {137}},\
  \bibinfo {pages} {A448} (\bibinfo {year} {1965})}\BibitemShut {NoStop}%
\bibitem [{\citenamefont {Latta}\ \emph {et~al.}(2009)\citenamefont {Latta},
  \citenamefont {H\"ogele}, \citenamefont {Zhao}, \citenamefont {Vamivakas},
  \citenamefont {Maletinsky}, \citenamefont {Kroner}, \citenamefont {Dreiser},
  \citenamefont {Carusotto}, \citenamefont {Badolato}, \citenamefont {Schuh},
  \citenamefont {Wegscheider}, \citenamefont {Atat\"ure},\ and\ \citenamefont
  {Imamoglu}}]{Latta2009}%
  \BibitemOpen
  \bibfield  {author} {\bibinfo {author} {\bibfnamefont {C.}~\bibnamefont
  {Latta}}, \bibinfo {author} {\bibfnamefont {A.}~\bibnamefont {H\"ogele}},
  \bibinfo {author} {\bibfnamefont {Y.}~\bibnamefont {Zhao}}, \bibinfo {author}
  {\bibfnamefont {A.~N.}\ \bibnamefont {Vamivakas}}, \bibinfo {author}
  {\bibfnamefont {P.}~\bibnamefont {Maletinsky}}, \bibinfo {author}
  {\bibfnamefont {M.}~\bibnamefont {Kroner}}, \bibinfo {author} {\bibfnamefont
  {J.}~\bibnamefont {Dreiser}}, \bibinfo {author} {\bibfnamefont
  {I.}~\bibnamefont {Carusotto}}, \bibinfo {author} {\bibfnamefont
  {A.}~\bibnamefont {Badolato}}, \bibinfo {author} {\bibfnamefont
  {D.}~\bibnamefont {Schuh}}, \bibinfo {author} {\bibfnamefont
  {W.}~\bibnamefont {Wegscheider}}, \bibinfo {author} {\bibfnamefont
  {M.}~\bibnamefont {Atat\"ure}}, \ and\ \bibinfo {author} {\bibfnamefont
  {A.}~\bibnamefont {Imamoglu}},\ }\href@noop {} {\bibfield  {journal}
  {\bibinfo  {journal} {Nature Physics}\ }\textbf {\bibinfo {volume} {5}},\
  \bibinfo {pages} {758} (\bibinfo {year} {2009})}\BibitemShut {NoStop}%
\bibitem [{\citenamefont {H\"ogele}\ \emph {et~al.}(2012)\citenamefont
  {H\"ogele}, \citenamefont {Kroner}, \citenamefont {Latta}, \citenamefont
  {Claassen}, \citenamefont {Carusotto}, \citenamefont {Bulutay},\ and\
  \citenamefont {Imamoglu}}]{Hogele2012}%
  \BibitemOpen
  \bibfield  {author} {\bibinfo {author} {\bibfnamefont {A.}~\bibnamefont
  {H\"ogele}}, \bibinfo {author} {\bibfnamefont {M.}~\bibnamefont {Kroner}},
  \bibinfo {author} {\bibfnamefont {C.}~\bibnamefont {Latta}}, \bibinfo
  {author} {\bibfnamefont {M.}~\bibnamefont {Claassen}}, \bibinfo {author}
  {\bibfnamefont {I.}~\bibnamefont {Carusotto}}, \bibinfo {author}
  {\bibfnamefont {C.}~\bibnamefont {Bulutay}}, \ and\ \bibinfo {author}
  {\bibfnamefont {A.}~\bibnamefont {Imamoglu}},\ }\href {\doibase
  10.1103/PhysRevLett.108.197403} {\bibfield  {journal} {\bibinfo  {journal}
  {Phys. Rev. Lett.}\ }\textbf {\bibinfo {volume} {108}},\ \bibinfo {pages}
  {197403} (\bibinfo {year} {2012})}\BibitemShut {NoStop}%
\bibitem [{\citenamefont {Seidl}\ \emph {et~al.}(2008)\citenamefont {Seidl},
  \citenamefont {Gerardot}, \citenamefont {Dalgarno}, \citenamefont {Kowalik},
  \citenamefont {Holleitner}, \citenamefont {Petroff}, \citenamefont {Karrai},\
  and\ \citenamefont {Warburton}}]{Seidl2008}%
  \BibitemOpen
  \bibfield  {author} {\bibinfo {author} {\bibfnamefont {S.}~\bibnamefont
  {Seidl}}, \bibinfo {author} {\bibfnamefont {B.~D.}\ \bibnamefont {Gerardot}},
  \bibinfo {author} {\bibfnamefont {P.~A.}\ \bibnamefont {Dalgarno}}, \bibinfo
  {author} {\bibfnamefont {K.}~\bibnamefont {Kowalik}}, \bibinfo {author}
  {\bibfnamefont {A.~W.}\ \bibnamefont {Holleitner}}, \bibinfo {author}
  {\bibfnamefont {P.~M.}\ \bibnamefont {Petroff}}, \bibinfo {author}
  {\bibfnamefont {K.}~\bibnamefont {Karrai}}, \ and\ \bibinfo {author}
  {\bibfnamefont {R.~J.}\ \bibnamefont {Warburton}},\ }\href@noop {} {\bibfield
   {journal} {\bibinfo  {journal} {Physica E}\ }\textbf {\bibinfo {volume}
  {40}},\ \bibinfo {pages} {2153} (\bibinfo {year} {2008})}\BibitemShut
  {NoStop}%
\end{thebibliography}
\end{document}